\documentclass[11pt]{article}
\usepackage{amssymb}

\usepackage{amsmath}
\usepackage{bm}
\usepackage{mathrsfs}
\usepackage{cite}
\begin{document}

\title{\textsc{Spin Magnetohydrodynamics}}
\author{G. Brodin\footnote{E-mail address: \texttt{gert.brodin@physics.umu.se}} $^\ddag$
  and M. Marklund\footnote{E-mail address: \texttt{mattias.marklund@physics.umu.se}} \footnote{Also at: Centre for Fundamental Physics, Rutherford Appleton Laboratory, Chilton, Didcot, Oxon OX11 OQX, U.K.}  \\
\textit{Department of Physics, Ume{\aa} University, SE--901 87 Ume{\aa},
Sweden}}
\date{{\small (Submitted to New J.\ Phys., December 22, 2006; Resubmitted June 14, 2007; 
Accepted July 20, 2007)}}
\maketitle

\begin{abstract}
Starting from the non-relativistic Pauli description of spin $\tfrac {1}{2}$
particles, a set of fluid equations, governing the dynamics of such
particles interacting with external fields and other particles, is derived.
The equations describe electrons, positrons, holes, and similar
conglomerates. In the case of electrons, the magnetohydrodynamic limit of an electron--ion plasma is
investigated. The results
should be of interest and relevance both to laboratory and astrophysical
plasmas. \\[3mm]
\end{abstract}


\section{Introduction}


The concept of a magnetoplasma has attracted interest ever since first
introduced by Alfv\'en \cite{alfven}, who showed the existence of waves in
magnetized plasmas. Since then, magnetohydrodynamics (MHD) has grown into a
vast and mature field of science, with applications ranging from solar
physics and astrophysical dynamos, to fusion plasmas and dusty laboratory
plasmas.

Meanwhile, a growing interest in what is known as quantum plasmas has
appeared (see, e.g., \cite{manfredi,marklund-brodin}). Here a main line of
research can be found starting from the Schr\"odinger description of the
electron. Assuming that the wave function can be factorized \footnote{%
There are thus no entanglement properties contained in the model.}, one may
derive a set of fluid equation for the electrons, starting either from an $N$%
-body description, a density matrix description, or a
Madelung (or Bohm) description of the wave function(s) \cite
{manfredi,holland}. As in classical fluid mechanics, the set of equationsn
may be closed by a suitable assumption concerning the thermodynamical
relation between quantities. These descriptions of the electron fluid, and
its interaction with ions and charged dust particles, has been shown to find
applications in many different settings \cite
{haas-etal1,anderson-etal,haas-etal2,haas,garcia-
etal,marklund,shukla-stenflo,shukla,Shukla-Eliasson,shukla-etal}. 
Part of the literature has been motivated by recent high field 
experimental progress and techniques \cite{marklund-shukla,exp1,exp2}, 
or the emergence of new areas, such as spintronics \cite{wolf-etal}.

Indeed, from the experimental perspective, a certain interest has been
directed towards the relation of spin properties to the classical theory of
motion (see, e.g., \cite
{halperin-hohenberg,balatsky,rathe-etal,hu-keitel,arvieu-etal,aldana-roso,walser-keitel,qian-vignale,walser-etal,roman- etal,liboff,fuchs-etal,kirsebom-etal}%
). In particular, the effects of strong fields on single particles with spin
has attracted experimental interest in the laser community \cite
{rathe-etal,hu-keitel,arvieu- etal,aldana-roso,walser-keitel,walser-etal}.
However, the main objective of these studies was single particle dynamics, 
relevant for dilute laboratory systems, whereas our focus will be on
collective effects. Moreover, strong
external magnetic fields can be found in astrophysical environments such as
pulsar \cite{Beskin-book,asseo} and magnetars \cite{magnetar}. Therefore, a
great deal of interest has been directed towards finding good descriptions of quantum
plasmas in such environments \cite
{melrose,melrose-weise,baring-etal,harding-lai}. Thus, there is ample need
and interest in developing models that are suitable for a wide range of
applications, taking into account collective effects in multi-particle
systems. There are in the literature several different studies of the effect of spin in plasma
systems, e.g.\ kinetic descriptions in connection to fusion studies \cite{cowley-etal,kulsrud-etal} and
electron spin waves in solid-state plasmas \cite{blum}. Moreover, the connection between 
microscopic and macroscopic spin dynamics in the plasma state has been discussed by 
de Groot and Suttorp \cite{degroot-suttorp}.

Inspired by both the historic and recent progress on quantum plasmas, a
complete set of multi-fluid spin plasma equations was presented in Ref.\ 
\cite{marklund-brodin}. In the current paper, we show, starting from the
non-relativistic Pauli equation for spin $\tfrac{1}{2}$ particles, how a
set of plasma equations can be derived for such spin $\tfrac{1}{2}$
particles. These particles may constitute electrons, positrons (albeit
non-relativistic), holes, or similar. Allowing these to interact with ions
or charged dust particles, as well as other spin $\tfrac{1}{2}$ particles,
gives the desired governing dynamics of spin plasmas. We furthermore derive
the appropriate magnetohydrodynamic description for such quantum plasmas,
and investigate the effects of spin on the dynamics of the plasma. 
The limitations and suitable parameter ranges of the derived governing equations
are discussed. The results should be of interest for both laboratory and
astrophysical plasmas.


\section{Governing equations}


In relativistic quantum mechanics, the spin of the electron (and positron) 
is rigorously introduced through the Dirac Hamiltonian 
\begin{equation}\label{eq:dirac}
  H = c\bm{\alpha}\cdot\left( \bm{p} + \frac{e}{c}\bm{A} \right) - e\phi + \beta m_ec^2 ,
\end{equation}
where $\bm{\alpha} = (\alpha_1, \alpha_2, \alpha_3)$, $e$ is the magnitude of the electron charge,
$c$ is the speed of light, $\bm{A}$ is the vector potential, $\phi$ is the electrostatic potential,
and the relevant matrices are given by
\begin{equation}
  \bm{\alpha} = \left( 
    \begin{array}{cc}
      0                 & \bm{\sigma} \\
      \bm{\sigma} & 0
    \end{array}
  \right) , \qquad
  {\beta} = \left( 
    \begin{array}{cc}
      {\mathsf{I}} & 0 \\
      0                     & -{\mathsf{I}}
    \end{array}
  \right) .
\end{equation}
Here ${\mathsf{I}}$ is the unit $2\times 2$ matrix and $\bm{\sigma} = (\sigma_1,\sigma_2, \sigma_3)$, where we have the Pauli spin matrices 
\begin{equation}
\sigma_1 = \left( 
\begin{array}{cc}
0 & 1 \\ 
1 & 0
\end{array}
\right) , \, \sigma_2 = \left( 
\begin{array}{cc}
0 & -i \\ 
i & 0
\end{array}
\right) , \, \text{ and }\, \sigma_3 = \left( 
\begin{array}{cc}
1 & 0 \\ 
0 & -1
\end{array}
\right) .
\end{equation}
From the Hamiltonian (\ref{eq:dirac}), a nonrelativistic counterpart may be obtained, 
taking the form
\begin{equation}\label{eq:pauli-hamiltonian}
  H = \frac{1}{2m_e}\left( \bm{p} + \frac{e}{c}\bm{A} \right)^2 
    + \frac{e\hbar}{2m_ec}\bm{B}\cdot\bm{\sigma} - e\phi .
\end{equation}
Thus, the electron possesses a magnetic moment $\bm{m} = 
-\mu_B\langle\psi|\bm{\sigma}|\psi\rangle/\langle\psi|\psi\rangle$, where 
$\mu_B = e\hbar/2m_ec$ is the Bohr magneton, giving a contribution $-\bm{B}\cdot\bm{m}$
to the energy. The latter shows the paramagnetic property of the electron, where the spin vector is
anti-parallel to the magnetic field in order to 
minimize the energy of the magnetized system.  
According to (\ref{eq:pauli-hamiltonian}) and the relation $dF/dt = \partial F/\partial t + (1/i\hbar)[F, H]$,
where $F$ is some operator and $[,]$ is the Poisson bracket, we have the following
evolution equations for the position and momentum in the Heisenberg picture
\cite{Dirac,Barut-Thacker}
\begin{equation}
  \bm{v} \equiv \frac{d\bm{x}}{dt} = \frac{1}{m_e}\left( \bm{p} + \frac{e}{c}\bm{A} \right) ,
\end{equation}
\begin{equation}
  m_e\frac{d\bm{v}}{dt} = -e\left( \bm{E} + \frac{\bm{v}}{c}\times\bm{B} \right) 
    - \frac{2}{\hbar}\mu_B\bm{\nabla}(\bm{B}\cdot\bm{S}) ,
\end{equation}
while the spin evolution is given by 
\begin{equation}
  \frac{d\bm{S}}{dt} = \frac{2}{\hbar}\mu_B \bm{B}\times\bm{S}, 
\end{equation}
where the spin operator is given by 
\begin{equation}
  \bm{S} = \frac{\hbar}{2}\bm{\sigma} .
\end{equation}
The above equations thus gives the quantum operator equivalents of the 
equations of motion for a classical particle, including the evolution of the
spin in a magnetic field. 

Next, we derive a set of nonlinear fluid equations for a gas of 
interacting electrons/positrons/holes (for a discussion of a kinetic approach, 
see Refs.\ \cite{cowley-etal} and \cite{kulsrud-etal}). 
The non-relativistic evolution of spin $\tfrac{1}{2}$ particles, as
described by the two-component spinor $\varPsi_{(\alpha)}$, is given by
(see, e.g.\ \cite{holland}) 
\begin{equation}  \label{eq:pauli}
i\hbar\frac{\partial\varPsi_{(\alpha)}}{\partial t} = \left[ -\frac
{\hbar^2}{2m_{j(\alpha)}}\left( \bm{\nabla} - \frac{iq_{(\alpha)}}{\hbar c}%
\bm{A} \right)^2 - \mu_{(\alpha)}\bm{B}\cdot\bm{\sigma} + q_{(\alpha)}\phi %
\right] \varPsi_{(\alpha)}
\end{equation}
where $m_{(\alpha)}$ is the particle mass, $\bm{A}$ is the vector potential, 
$q_{(\alpha)}$ the particle charge, $\mu_{(\alpha)}$ the particle's magnetic
moment, $\bm{\sigma} = (\sigma_1, \sigma_2, \sigma_3)$ the Pauli spin
matrices, $\phi$ the electrostatic potential, and ${(\alpha)}$ enumerates
the wave functions. For an electron, the magnetic moment is given by $\mu_e = -\mu_B
\equiv -e\hbar/2m_ec$. From now on, we will define $\mu_{(\alpha)} \equiv
\mu \equiv q\hbar/2mc$.

Next we introduce the decomposition of the spinors as 
\begin{equation}  \label{eq:decomp}
\varPsi_{(\alpha)} = \sqrt{ n_{(\alpha)}}\,\exp(iS_{(\alpha)}/\hbar)%
\varphi_{(\alpha)} ,
\end{equation}
where $n_{(\alpha)}$ is the density, $S_{(\alpha)}$ is the phase, and $%
\varphi_{(\alpha)}$ is the 2-spinor through which the spin $\tfrac{1} {2}$
properties are mediated. Multiplying the Pauli equation (\ref{eq:pauli}) by $%
\varPsi_{(\alpha)}^{\dag}$, inserting the decomposition (\ref{eq:decomp}),
and taking the gradient of the resulting phase evolution equation, we obtain
the continuity and moment conservation equation 
\begin{equation}  \label{eq:cont-alpha}
\frac{\partial n_{(\alpha)}}{\partial t} + \bm{\nabla}\cdot(n_ {(\alpha)}%
\bm{v}_{(\alpha)}) = 0
\end{equation}
and 
\begin{eqnarray}  \label{eq:mom-alpha}
&&\!\!\!\!\!\!\!\!\! m_{(\alpha)}\left( \frac{\partial}{\partial t} + \bm{v}%
_{(\alpha)}\cdot\bm{\nabla} \right)\bm{v}_{(\alpha)} = q_{(\alpha)} (\bm{E}
+ \bm{v}_{(\alpha)}\times\bm{B})  \notag \\
&&\!\!\!\!\!\!\!\!\! \quad + \frac{2\mu}{\hbar}(\bm{\nabla}\otimes\bm {B}%
)\cdot\bm{s}_{(\alpha)} - \bm{\nabla}Q_{(\alpha)} - \frac{1}{%
m_{(\alpha)}n_{(\alpha)}}\bm{\nabla}\cdot\left(n_{(\alpha)}%
\bm{\mathsf{\Sigma}}_{(\alpha)} \right)
\end{eqnarray}
respectively. The spin contribution to Eq.\ (\ref{eq:mom-alpha}) is consistent with the results
of Ref.\ \cite{degroot-suttorp}. Here we have introduced the tensor index $a, b, \ldots = 1, 2,
3$, the velocity is defined by 
\begin{equation}
\bm{v}_{(\alpha)} = \frac{1}{m_{(\alpha)}}\left( \bm{\nabla}S_ {(\alpha)} -
i\hbar\varphi^{\dag}\bm{\nabla}\varphi \right) - \frac{q_{(\alpha)}}{%
m_{(\alpha)}c}\bm{A} ,
\end{equation}
the Schr\"odinger like quantum potential (or Bohm potential) is given by 
\begin{equation}
Q_{(\alpha)} = -\frac{\hbar^2}{2m_{(\alpha)}n_{(\alpha)}^{1/2}}%
\nabla^2n_{(\alpha)}^{1/2} ,
\end{equation}
the spin density vector is 
\begin{equation}
\bm{s}_{(\alpha)} = \frac{\hbar}{2}\varphi_{(\alpha)}^{\dag}\bm{\sigma}%
\varphi_{(\alpha)} ,
\end{equation}
which satisfies $|\bm{s}_{(\alpha)}| = \hbar/2$, and we have defined the
symmetric gradient spin tensor 
\begin{equation}
\bm{\mathsf{\Sigma}}_{(\alpha)} = (\bm{\nabla}{s}_{(\alpha)a})\otimes(%
\bm{\nabla}{s}_{(\alpha)}^a) .
\end{equation}
Moreover, contracting Eq.\ (\ref{eq:pauli}) by $\varPsi_{(\alpha)}^ {\dag}%
\bm{\sigma}$, we obtain the spin evolution equation 
\begin{eqnarray}  \label{eq:spin-alpha}
\left( \frac{\partial}{\partial t} + \bm{v}_{(\alpha)}\cdot\bm{\nabla}
\right)\bm{s}_{(\alpha)} = -\frac{2\mu}{\hbar}\bm{B}\times\bm{s}_{{%
(\alpha)}} + \frac{1}{m_{(\alpha)}n_{(\alpha)}}\bm{s}_{{(\alpha)}} \times%
\left[\partial_a(n_{(\alpha)}\partial^a\bm{s}_{{(\alpha)}}) \right].
\end{eqnarray}
We note that the particles are coupled via Maxwell's equations.

Suppose that we have $N$ wave functions for the same particle species with
mass $m$, magnetic moment $\mu$, and charge $q$, and that the total system
wave function can be described by the factorization $\varPsi = \varPsi_{(1)}%
\varPsi_{(2)} \ldots \varPsi_{(N)}$. Then we define the total particle
density for the species with charge $q$ according to 
\begin{equation}  \label{eq:totdensity}
n_q = \sum_{{(\alpha)} = 1}^Np_{\alpha}n_{(\alpha)} ,
\end{equation}
where $p_{\alpha}$ is the probability related to the wave function $\varPsi%
_{(\alpha)}$. Using the ensemble average $\langle f\rangle =
\sum_{\alpha}p_{\alpha}(n_{(\alpha)}/n_q)f$ (for any tensorial quantity $f$),
the total fluid velocity for charges $q$ is $\bm{V}_q = \langle\bm{v}%
_{(\alpha)}\rangle$ and the total spin density is $\bm{S} = \langle\bm{s}%
_{(\alpha)}\rangle $. From these definitions we can define the microscopic
velocity in the fluid rest frame according to $\bm{w}_{(\alpha)} = \bm {v}%
_{(\alpha)} - \bm{V}$, satisfying $\langle\bm{w}_{(\alpha)}\rangle = 0 $,
and the microscopic spin density $\bm{\mathcal{S}}_{(\alpha)} = \bm{s}%
_{(\alpha)} - \bm{S}$, such that $\langle\bm{\mathcal{S}}_{(\alpha)} \rangle
= 0$.

Taking the ensemble average of Eqs.\ (\ref{eq:cont-alpha}), (\ref
{eq:mom-alpha}), and (\ref{eq:spin-alpha}), we obtain 
\begin{equation}
\frac{\partial n_{q}}{\partial t}+\bm{\nabla}\cdot (n_{q}\bm{V}_{q})=0,
\label{eq:density}
\end{equation}
\begin{equation}
mn_{q}\left( \frac{\partial }{\partial t}+\bm{V}_{q}\cdot \bm{\nabla}\right) %
\bm{V}_{q}=qn_{q}\left( \bm{E}+\bm{V}_{q}\times \bm{B}\right) -\bm {\nabla}%
\cdot \bm{\mathsf{\Pi}}_{q}-\bm{\nabla}P_{q}+\bm{\mathcal{C}}_{qi}+\bm {F}%
_{Q}  \label{eq:mom-q}
\end{equation}
and 
\begin{equation}
n_{q}\left( \frac{\partial }{\partial t}+\bm{V}_{q}\cdot \bm{\nabla}\right) %
\bm{S}= -\frac{2\mu n_{q}}{\hbar }\bm{B}\times \bm{S}-\bm{\nabla}\cdot {%
\bm{\mathsf{K}}}_{q}+\bm{\Omega}_{S}  \label{eq:spin-q}
\end{equation}
respectively. Here we have added the collisions $\bm{\mathcal{C}}_{qi}$
between charges $q$ and the ions $i$, denoted the total quantum force
density by 
\begin{eqnarray}
&&\bm{F}_{Q} = \frac{2\mu n_{q}}{\hbar }(\bm{\nabla}\otimes \bm{B})\cdot %
\bm{S}-n_{q}\langle \bm{\nabla}Q_{(\alpha )}\rangle -\frac{1}{m}\bm {\nabla}%
\cdot \left( n_{q}\bm{\mathsf{\Sigma}}\,\right) -\frac{1}{m}\bm {\nabla}%
\cdot \big(n_{q}\widetilde{\bm{\mathsf{\Sigma}}}\,\big)  \notag \\
&&\qquad \qquad -\frac{1}{m}\bm{\nabla}\cdot \big[n_{q}(\bm{\nabla}%
S_{a})\otimes \langle (\bm{\nabla}\mathcal{S}_{(\alpha )}^{a})\rangle
+n_{q}\langle (\bm{\nabla}\mathcal{S}_{(\alpha )a})\rangle \otimes (%
\bm{\nabla}S^{a})\big],
\end{eqnarray}
consistent with the results in Ref.\ \cite{degroot-suttorp}, and defined the the nonlinear spin fluid correction according to 
\begin{eqnarray}
&&\bm{\Omega}_{S}=\frac{1}{m}\bm{S}\times \lbrack \partial
_{a}(n_{q}\partial ^{a}\bm{S})]+\frac{1}{m}\bm{S}\times \lbrack \partial
_{a}(n_{q}\langle \partial ^{a}\bm{\mathcal{S}}_{(\alpha )}\rangle )]  \notag
\\
&&\qquad +\frac{n_{q}}{m}\left\langle \frac{\bm{\mathcal{S}}_{(\alpha )}}{%
n_{(\alpha )}}\times \lbrack \partial _{a}(n_{(\alpha )}\right\rangle
\partial ^{a}\bm{{S}})]+\frac{n_{q}}{m}\left\langle \frac{\bm{\mathcal {S}}%
_{(\alpha )}}{n_{(\alpha )}}\times \lbrack \partial _{a}(n_{(\alpha
)}\partial ^{a}\bm{\mathcal{S}}_{(\alpha )})]\right\rangle ,
\end{eqnarray}
where $\bm{\mathsf{\Pi}}=mn[\langle \bm{w}_{(\alpha )}\otimes \bm{w}%
_{(\alpha )}\rangle -\bm{\mathsf{I}}\langle w_{(\alpha )}^{2}\rangle /3]$ is
the trace-free anisotropic pressure tensor ($\bm{\mathsf{I}}$ is the unit
tensor), $P=mn\langle w_{(\alpha )}^{2}\rangle $ is the isotropic scalar
pressure, $\bm{\mathsf{\Sigma}}=(\bm{\nabla}S_{a})\otimes (\bm{\nabla}S^{a})$
is the nonlinear spin correction to the classical momentum equation, $%
\widetilde{\bm{\mathsf{\Sigma}}}=\langle (\bm{\nabla}\mathcal{S}_{(\alpha
)a})\otimes (\bm{\nabla}\mathcal{S}_{(\alpha )}^{a})\rangle $ is a pressure
like spin term (which may be decomposed into trace-free part and trace), ${%
\bm{\mathsf{K}}}=n\langle \bm{w}_{(\alpha )}\otimes \bm{\mathcal{S}}%
_{(\alpha )}\rangle $ is the thermal-spin coupling, and $[(\bm{\nabla}%
\otimes \bm{B})\cdot \bm{S}\,]^{a}=(\partial ^{a}B_{b})S^{b}$. 
Here the indices $a,b,\ldots = 1,2,3$ denotes the Cartesian components 
of the corresponding tensor.
We note that
the momentum conservation equation (\ref{eq:mom-q}) and the spin evolution
equation (\ref{eq:spin-q}) still contains the explicit sum over the $N$
states. 

The coupling between the quantum plasma species is mediated by the
electromagnetic field. By definition, we let $\bm{H} = \bm{B}/\mu_0 - \bm{M}$
where $\bm{M} = 2n_q\mu\bm{S}/\hbar$ is the magnetization due to the spin sources. 
Amp\`ere's law $\bm{\nabla}\times\bm{H} = \bm{j} + \epsilon_0\partial_t\bm{E}$ 
then takes the form 
\begin{equation}
  \bm{\nabla} \times \bm{B}=\mu _{0}(\bm{j}+\bm{j}_{M}) 
    + \frac{1}{c^2}\frac{\partial\bm{E}}{\partial t},  \label{Eq-ampere}
\end{equation}
where we have the magnetization spin current $\bm{j}_{M}= \bm{\nabla} \times
\bm{M}$ and the free current $\bm{j}$. 
The system is closed by Faraday's law 
\begin{equation}
  \bm{\nabla} \times \bm{E}=-\frac{\partial\bm{B}}{\partial t} . \label{Eq-Faraday}
\end{equation}


\section{Electron--ion plasma and the magnetohydrodynamic limit}


The preceding analysis applies equally well to electrons as holes or similar
condensations. We will now assume that the quantum particles are electrons,
thus $q=-e$, where $e$ is the magnitude of the electron charge. By the
inclusion if the ion species, which are assumed to be described by the
classical equations and have charge $Ze$, we may derive a set of one- fluid
equations. The ion equations read 
\begin{equation}
\frac{\partial n_{i}}{\partial t}+\bm{\nabla}\cdot (n_{i}\bm{V}_{i})=0,
\label{eq:ion-density}
\end{equation}
and 
\begin{equation}
m_{i}n_{i}\left( \frac{\partial }{\partial t}+\bm{V}_{i}\cdot \bm {\nabla}%
\right) \bm{V}_{i}=Zen_{i}\left( \bm{E}+\bm{V}_{i}\times \bm{B} \right) -%
\bm{\nabla}\cdot \bm{\mathsf{\Pi}}_{i}-\bm{\nabla}P_{i}+\bm{\mathcal {C}}%
_{iq}.  \label{eq:ion-mom}
\end{equation}
Next we define the total mass density $\rho \equiv (m_{e}n_{e}+m_{i}n_ {i})$%
, the centre-of-mass fluid flow velocity $\bm{V}\equiv (m_{e}n_{e}\bm{V}%
_{e}+m_{i}n_{i}\bm{V}_{i})/\rho $, and the current density $\bm{j}=- en_{e}%
\bm{V}_{e}+Zen_{i}\bm{V}_{i}$. Using these denfinitions, we immediately
obtain 
\begin{equation}
\frac{\partial \rho }{\partial t}+\bm{\nabla}\cdot (\rho \bm{V})=0,
\label{eq:mhd-cont}
\end{equation}
from Eqs.\ (\ref{eq:density}) and (\ref{eq:ion-density}). Assuming
quasi-neutrality, i.e.\ $n_{e}\approx Zn_{i}$, the momentum conservation
equations (\ref{eq:mom-q}) and (\ref{eq:ion-mom}) give 
\begin{equation}
\rho \left( \frac{\partial }{\partial t}+\bm{V}\cdot \bm{\nabla} \right) %
\bm{V}=\bm{j}\times \bm{B}-\bm{\nabla}\cdot \bm{\mathsf{\Pi}}-\bm {\nabla}P+%
\bm{F}_{Q},  \label{eq:mhd-mom}
\end{equation}
where $\bm{\mathsf{\Pi}}$ is the tracefree pressure tensor in the
centre-of-mass frame, $P$ is the scalar pressure in the centre-of-mass
frame, and the collisional contributions cancel due to momentum
conservation. We also note that due to quasi-neutrality, we have $n_
{e}=\rho /(m_{e}+m_{i}/Z)$ and $\bm{V}_{e}=\bm{V}-m_{i}\bm{j}/Ze\rho $, and
we can thus express the quantum terms in terms of the total mass density $%
\rho $, the centre-of-mass fluid velocity $\bm{V}$, and the current $\bm{j}$%
. With this, the spin transport equation (\ref{eq:spin-q}) reads 
\begin{equation}
\rho \left( \frac{\partial }{\partial t}+\bm{V}\cdot \bm{\nabla} \right) %
\bm{S}=\frac{m_{e}}{Ze}\bm{j}\cdot \bm{\nabla}\bm{S}-\frac{2\mu \rho } {%
\hbar }\bm{B}\times \bm{S}-\left( m_{e}+\frac{m_{i}}{Z}\right) \bm{\nabla}
\cdot {\bm{\mathsf{K}}}_{q}+\left( m_{e}+\frac{m_{i}}{Z}\right) \bm{\Omega}_
{S}.  \label{eq:mhd-spin}
\end{equation}

In the momentum equation (\ref{eq:mhd-mom}), we have the force density $\bm{j}\times\bm{B} + \bm{F}_Q$, where
$\bm{j}$ is the current produced by the free charges. In general, for a magnetized medium
with magnetization density $\bm{M}$, Amp\`ere's law gives the free current in a finite volume $V$ 
according to 
\begin{equation}\label{eq:free-current}
  \bm{j} = \frac{1}{\mu_0}\bm{\nabla}\times\bm{B} - \bm{\nabla}\times\bm{M}  ,
\end{equation}
where we have neglected the displacement current is $\bm{j}_D = \epsilon_0\partial_t\bm{E}$. The delta-function
contribution of the surface current is an important part of the total current when we are interested
in the forces on a finite volume, as will now be shown. 

It it worth noting that the expression of the force density in the momentum conservation equation 
can, to lowest order in the spin, be derived on general macroscopic grounds. 
Formally, the total force density on a 
volume element $V$ is defined as $\bm{F} 
= \lim_{V \rightarrow 0}(\sum_{\alpha} \bm{f}_{\alpha}/V)$,
where $\bm{f}_{\alpha}$ are the different forces acting on the volume element, and might
include surface forces as well. For 
magnetized matter, the total force on an element of volume $V$ is then
\begin{equation}
  \bm{f}_{\rm tot} = \int_V\bm{j}_{\rm tot}\times\bm{B}\,\mathrm{d}V 
    + \oint_{\partial V}(\bm{M}\times\hat{\bm{n}})\times\bm{B}\,\mathrm{d}S 
\end{equation}
where (neglecting the displacement current) $\bm{j}_{\rm tot} = \bm{j} + \bm{\nabla}\times\bm{M}$.
Inserting the expression for the total current into the volume integral and using the divergence
theorem on the surface integral, we obtain the force density
\begin{equation}
  \bm{F}_{\rm tot} = \bm{j}\times\bm{B}
    + M_k\bm{\nabla}B^k  ,
\end{equation}
identical to the lowest order description from the Pauli equation (see Eq.\ (\ref{eq:mhd-mom})). 
Inserting the free current expression (\ref{eq:free-current}), due to Amp\`ere's law, we can write
the total force density according to
\begin{equation}\label{eq:total-force}
   {F}^i = -\partial^i\left( 
    \frac{B^2}{2\mu_0} - \bm{M}\cdot\bm{B} 
  \right) + \partial_k({H^iB^k}) .
\end{equation}
The first gradient term in Eq.\ (\ref{eq:total-force}) can be interpreted as the 
force due to a potential (the energy of the magnetic field and the magnetization
vector in that field), while the second divergence term is the anisotropic magnetic pressure effect. 
Noting that the spatial part of the stress tensor takes the form
$T^{ik} = -H^iB^k +(B^2/2\mu_0 - \bm{M}\cdot\bm{B} )\delta^{ik}$  \cite{degroot-suttorp}, 
we see that the
total force density on the magnetized fluid element can be written $F^i = -\partial_kT^{ik}$, 
as expected. Thus, the Pauli theory results in the same type of conservation laws as
the macroscopic theory. 
The momentum conservation equation (\ref{eq:mhd-mom}) then reads
\begin{equation}
  \rho \left( \frac{\partial }{\partial t}+\bm{V}\cdot \bm{\nabla} \right)\bm{V} = %
  -\bm{\nabla}\left( 
    \frac{B^2}{2\mu_0} - \bm{M}\cdot\bm{B} 
  \right) + B^k\partial_k\bm{H}
  -\bm {\nabla}P
  ,  \label{eq:mhd-mom2}
\end{equation}
where for the sake of clarity we have assumed an isotropic pressure, dropped the displacement
current term in accordance with the nonrelativistic assumption, and neglected the
Bohm potential (these terms can of course simply be added to (\ref{eq:mhd-mom2})).

Approximating the quantum corrections, using $L\gg \lambda
_{F}$ where $L$ is the typical fluid length scale and $\lambda _{F}$ is the
Fermi wavelength for the particles, according to \cite{manfredi} 
\begin{equation}
\langle \bm{\nabla}Q_{(\alpha )}\rangle \approx -\bm{\nabla}\left( \frac{%
\hbar ^{2}}{2mn_q^{1/2}}\nabla ^{2}n_q^{1/2}\right) \equiv \bm{\nabla}Q.
\end{equation}
We then note that even if $Q$ is small, the magnetic field may, through the
dynamo equation (\ref{Eq-Faraday}), still be driven by pure quantum effects
through the spin.

A generalized Ohm's law may be derived assuming $\bm{\mathcal{C}}%
_{ei}=en_{e}\eta \bm{j}$, where $\eta $ is the resistivity. From the
electron mometum conservation equation (\ref{eq:mom-q}) combined with
Faraday's law we obtain 
\begin{equation}
\frac{\partial \bm{B}}{\partial t}=\bm{\nabla}\times \left\{ \bm{V} \times %
\bm{ B}-\frac{\bm{j}\times \bm{B}}{%
en_e}-\eta \bm{j}+\frac{m_{e}}{e}\frac{d\bm{V}_{e}}{dt}-\frac {\bm{F}%
_{Q}}{en_e}\right\}  \label{eq:ohm1}
\end{equation}
where $\eta $ is the resistivity. Here we have omitted the anisotropic part
of the pressure, and neglected terms of order $m_{e}/m_{i}$ compared with
unity.

The electron inertia term is negligible unless the electron velocity is much
larger than the ion velocity. Thus whenever electron inertia is important,
we include only the electron contribution to the current, and use Amp\`ere's
law to substitute $\bm{V}_{e}=\bm{\nabla}\times \bm{B/}en_e\mu _{0}$ into the
term proportional to $m_{e}$ in (\ref{eq:ohm1}), which gives the final form
of the Generalized Ohm's law 
\begin{equation}
\frac{\partial \bm{B}}{\partial t}=\bm{\nabla}\times \Bigg\{\bm{V}\times %
\bm{B}-\frac{\bm{j}\times \bm {B}}{%
en_e}-\eta \bm{j}-\frac{m_{e}}{e^{2}\mu _{0}}\left[ \frac{\partial }{%
\partial t}-\left( \frac{\bm{\nabla}\times \bm{B}}{e\mu _{0}n_e}\right) \cdot %
\bm{\nabla}\right] \frac{\bm{\nabla}\times \bm{B}}{n_e}-\frac{\bm{F}_{Q}}{en_e}%
\Bigg\}  \label{eq:dynamo}
\end{equation}

In the standard MHD regime the Hall term and the electron inertia term are
negligible. During such conditions the quantum force is also negligible in
Ohm's law, which reduces to its standard MHD form 
\begin{equation}
\frac{\partial \bm{B}}{\partial t}=\bm{\nabla}\times \left( \bm{V}\times %
\bm{B}-\eta \bm{j}\right) 
\end{equation}
Note, however, that the quantum force including the spin effects still
should be kept in the momentum equation (\ref{eq:mhd-mom}). Equations
 (\ref{eq:mhd-cont}),  (\ref
{eq:mhd-mom}), and (\ref{eq:dynamo}) together with the
spin evolution equation (\ref{eq:mhd-spin}), which is needed to determine $%
\bm{F}_{Q}$, constitutes the basic
equations. In order to close the system, equations of state for the pressure
as well as for the spin state are needed, as will be discussed in section 5.


\section{Closing the system}


The momentum equation, Ohm's (generalized) law and the continuity equation
need to be completed by an equation of state for the pressure. As is
well-known, rigorous derivations of the equation of state is only applicable
in special cases of limited applicability to real plasmas. Useful models
include a scalar pressure where the pressure is proportional to some power
of the density, i.e. 
\begin{equation}
\frac{d}{dt}\left( \frac{P_{s}}{n_{s}^{\gamma _{s}}}\right) =0 ,\label{Eq-state}
\end{equation}
where $d/dt=\partial /\partial t+\bm{V}_{s}\cdot \nabla $ is the
convective derivative and $\gamma _{s}$ is the ratio of specific heats,
which in general can be different for different species $s$. Secondly, we
note that the magnitude of the terms that are quadratic in the spin depends
highly on the spatial scale of the variations. In MHD, the scale lengths are
typically longer or equal to the Larmor radius of the heavier particles, which
means that the terms that are quadratic in $S$ can be neglected in the
expression for the quantum force $\bm{F}_{Q}$ as well as in the spin
evolution equation (\ref{eq:mhd-spin}). To lowest order, the spin inertia
can be neglected for frequencies well below the electron cyclotron
frequency. Also omitting the spin-thermal coupling term, which is small for
the same reasons as stated above, the spin-vector is determined from 
\begin{equation}
\bm{B}\times \bm{S}=0.
\end{equation}
which has a solution 
\begin{equation}
\bm{S} = -\frac{\hbar }{2}\,\eta \!\left( \frac{\mu _{B}B}{k_{B}T_{e}}\right) 
\widehat{\bm{B}}  \label{eq:spin-closing}
\end{equation}
consistent with standard theories of paramagnetism, 
as the spin anti-parallel to the magnetic field minimizes the magnetic
moment energy, so that 
\begin{equation}
  \bm{M} = 2\mu_Bn_e\,\eta\!\left(\frac{\mu_BB}{k_BT_e}\right)\widehat{\bm{B}} .
\end{equation} 
Here $B$ denotes the magnitude 
of the magnetic field, $\eta (x) =\tanh x$ is the
Brillouin function,\footnote{%
  In general, the thermodynamic equlibrium distribution of spin results in a magnetization proportional to the Brillouin function $B_j(\mu_B/k_B T)$, where $j$ is the spin of the particles in question.
For the special case of spin $1/2$-particles we have $B_{1/2}(x)=\tanh(x)$.
} 
and $\widehat{\bm{B}}$ is a unit vector in the direction of the magnetic
field. In this approximation the spin evolution equation (\ref{eq:mhd-spin})
can be dropped, and the quantum force can be written 
\begin{equation}
\bm{F}_{Q} = n_{e}\bm{\nabla}\left( \frac{\hbar ^{2}}{2mn_{e}^{1/2}}\nabla
^{2}n_{e}^{1/2}\right) + 2n_e\,\eta \!\left( \frac{\mu _{B}B}{k_{B}T_{e}}\right)
\mu_B \bm{\nabla} B  \label{eq:f-closed}
\end{equation}
where the second term comes from the spin. Combining the approximations
presented in this section together with the MHD equations presented in
section 3, or either of the two dust systems presented in section 4,
closed systems are obtained.


\section{Illustrative example}


Let us next consider the spin-model described by (28), (35), and (39) with the
resistivity put to zero. Furthermore, we note that for most plasmas $\mu
_{B}B\ll k_{B}T$, and thus we can use the approximation $\eta (\mu
_{B}B/k_{B}T)\approx \mu _{B}B/k_{B}T$. For definiteness, we also choose an
isothermal equation of state such that $\bm{\nabla}P=k_BT\bm{\nabla} n$. Next we let $%
\bm{B} = B_{0}\widehat{\bm{z}} + \bm{B}_{1}$, $\bm{M} = M_{0}%
\widehat{\bm{z}} + \bm{M}_{1}$, $n = n_{0} + n_{1},$ where index $0$
denotes the equilibrium part and index $1$ denote the perturbation, and we
have assumed that the equilbrium part of the velocity is zero. Linearizing
around the equilibrium and Fourier analysing, and omitting the non-spin part
of the quantum force,\footnote{%
  The ordinary part of the quantum force is negligible
compared to the spin coupling provided $eB_{0}\gg \hbar k^{2}$.
} we find that the general dispersion
relation can be written
\begin{equation}
\left( \omega ^{2}-k^{2}\widetilde{C}_{A}^{2}\right) \left[ \left( \omega
^{2}-k^{2}\widetilde{C}_{A}^{2}+k_{x}^{2}\widetilde{c}_{s}^{2}\right)
(\omega ^{2}-k_{z}^{2}c_{s}^{2})+k_{x}^{2}k_{z}^{2}\widetilde{c}_{s}^{4}%
\right] =0  \label{MHD-dispersion}
\end{equation}
where  $c_{s}=[(k_{B}T_{e}+k_{B}T_{i})/m_{i}]^{1/2}$ is the standard
acoustic velocity. Formally, Eq. (\ref{MHD-dispersion}) looks similar to the
standard ideal MHD dispersion relation. However, we note that firstly spin
effects appear in the spin-modified Alfv\'{e}n velocity  \begin{equation}
\widetilde{C}_{A}=C_{A}\left( 1-\frac{\hbar ^{2}\omega _{pe}^{2}}{%
mc^{2}k_{B}T}\right) ^{1/2}  \label{Eq-cA-mod}
\end{equation}
where $\omega _{pe}$ is the electron plasma frequency and $C_{A}=\left(
B_{0}^{2}/\mu _{0}\rho _{0}\right) ^{1/2}$ is the standard Alfv\'{e}n
velocity. Moreover, another spin-modification is introduced in the factor $%
\widetilde{c}_{s}$, given by
\begin{equation}
\widetilde{c}_{s}=c_{s}\left[ 1-\left( \frac{\hbar \omega _{ce}}{k_{B}T}%
\right) ^{^{2}}\right] ^{1/2}  \label{Eq-cs-mod}
\end{equation}
The first factor of Eq. (\ref{MHD-dispersion})  describes the spin-modified
shear Alfv\'{e}n waves, and the second factor describes fast and slow
magnetosonic waves. We point out that the assumption made in this
section initially, $\mu _{B}B\ll k_{B}T$, implies that both $\widetilde{C}%
_{A}$ and $\widetilde{c}_{s}$ given by (\ref{Eq-cA-mod}) and (\ref{Eq-cs-mod}%
) respectively are real. Moreover, we stress that we cannot
simply obtain the spin-modified magnetosonic dispersion relation from the
classical one by the substitutions $C_{A}\rightarrow \widetilde{C}_{A}$ and $%
c_{s}\rightarrow \widetilde{c}_{s}$, since the classical acoustic velocity $%
c_{s}$ still appears in the factor $(\omega ^{2}-k_{z}^{2}c_{s}^{2})$, as
seen from (\ref{MHD-dispersion}). We end this section by noting that the
modification of the Alfv\'{e}n velocity can be appreciable for a high
density low temperature plasma.


\section{Summary and Discussion}


In the present paper we have derived one-Fluid MHD equations for a number of
different plasmas including the effects of the electron spin, starting from
the Pauli equation for the individual particles. In particular we have
derived spin-MHD equations for an electron-ion plasma. 
Furthermore, the general equations derived in section 2 constitutes a
basis for an electron--positron plasma description including spin effects.

In order to obtain closure of the system, our equations needs to be supplemented by
equations of state for the pressure as well as for the spin pressure. In the
MHD regime, a rather simple way to achieve this closure has been discussed
in Section 4, where we assume that the scale lengths are long enough such
terms that are quadratic in the spin vector as well as the spin-thermal
coupling are neglected. However, we here note that if more elaborate models
are used, the spin pressure together with the spin-thermal coupling might
play an important role in the generalized Ohms's law (\ref{eq:dynamo}).

Since the spin-coupling give raise to a parallel force (to the magnetic
field) in the momentum equation, the parallel electric field will not be
completely shielded even for zero temperature, contrary to ordinary MHD. As an
immediate consequence, the spin-coupling can give rise to a rich variety of
physical effects. In this paper we have limited ourselves to present a single
example, linear wave propagation in an electron--ion plasma, and shown the
modification of the dispersion relation due to the spin effects. In
particular, we note that the spin effects are important for low temperature, high densities, and/or
strongly magnetized plasmas. The latter can be found in astrophysical systems, such as
pulsars or magnetars \cite{harding-lai}, where however other modifications, such as vacuum 
polarization and magnetization \cite{marklund-shukla,brodin-etal}, of our set
of governing equations may be necessary in order to accurately model such plasmas. 
Moreover, ultra-cold plasmas may even be found in 
laboratory environments, where currently mK temperature plasmas can be
formed \cite{Robinson-etal}. Studies involving the spin dynamics through the
spin evolution equation (\ref{eq:spin-q}), kinetic effects associated with
the spin, as well as nonlinear spin dynamics are projects for future work.
Finally, we note that dusty plasmas can sustain weakly
damped modes with low phase velocities \cite{shukla-mamun}, and quantum and
spin effects tend to be important in this regime.

\subsubsection*{Acknowledgments}
We would like to thank Padma K. Shukla for interesting and fruitful 
discussions. This research was supported by the Swedish Research Council.


\end{document}